\documentclass[epj]{webofc}
\usepackage[varg]{txfonts}   
\usepackage{graphicx,amsmath,wrapfig,epsfig}
\woctitle{XLVI International Symposium on Multiparticle Dynamics}

\begin{document}
\title{Hadron tomography studies by generalized parton distributions 
and distribution amplitudes}
\author{S. Kumano\inst{1,2}}
\institute{
   KEK Theory Center, Institute of Particle and Nuclear Studies, KEK \\
   \ \ and Department of Particle and Nuclear Physics,
   Graduate University for Advanced Studies, \\
   \ \ 1-1, Ooho, Tsukuba, Ibaraki, 305-0801, Japan 
\and
   J-PARC Branch, KEK Theory Center,
   Institute of Particle and Nuclear Studies, KEK \\
   \ \ and Theory Group, Particle and Nuclear Physics Division, 
   J-PARC Center, \\
   \ \ 203-1, Shirakata, Tokai, Ibaraki, 319-1106, Japan
}
\abstract{
We discuss hadron-tomography studies for the nucleon and exotic hadrons
by high-energy hadron reactions. First, the constituent-counting rule 
is explained for determining internal quark configurations of 
exotic-hadron candidates by scaling properties of high-energy exclusive 
cross sections. Next, possibilities are discussed for investigating 
the generalized parton distributions (GPDs) of the nucleon and 
exotic hadrons at J-PARC. In particular, we study hadronic $2 \to 3$ 
process $p+p \to N+\pi+B$, exclusive Drell-Yan process, and 
exotic-hadron GPDs. For determining three-dimensional structure of 
unstable exotic hadrons, we consider $s$-$t$ crossed quantities of 
the GPDs called generalized distribution amplitudes (GDAs), 
which can be investigated at KEKB.
We explain possible studies of the GDAs by two-photon processes.
}

\maketitle

\section{Introduction}
\label{intro}

Internal structure of the nucleon has been investigated for a long time
by lepton and hadron scattering processes at various energies,
and major properties are now understood. However, the origin of 
the nucleon spin has not been clarified yet although it is
one of fundamental physics quantities. For understanding the whole picture
of the nucleon spin in terms of quarks and gluons, it is essential
to understand partonic orbital-angular-momentum contributions.
Therefore, it became necessary to establish three-dimensional (3D)
picture of the nucleon, and this field is called hadron tomography.
It is now under active investigations through
the 3D structure functions of generalized parton 
distributions (GPDs) and transverse-momentum-dependent parton 
distributions (TMDs). There are also generalized distribution
amplitudes (GDAs) which are $s$-$t$ crossed quantities to the GPDs.

These 3D structure functions are key quantities
in solving the nucleon-spin puzzle. However, it is interesting
to use them for exotic-hadron studies because it is not easy
to find internal quark-gluon configurations,
namely undoubted evidence of exotic nature, 
solely by low-energy observables such as masses and decay widths. 
In general, high-energy reactions should be useful for finding
internal structure of exotic hadrons by using, for example,
fragmentation functions \cite{ffs-exotic}, 
constituent-counting rule \cite{kks2013,cks2016}, and
the 3D structure functions \cite{gpd-gda}
because the appropriate degrees of freedom are quark and gluons.

In this article, possible exotic-hadron studies are proposed 
by using high-energy reactions, together with the usual nucleon GPDs,
at facilities such as J-PARC and KEKB. 
First, we explain the constituent-counting
rule, predicted by perturbative QCD, as a useful guidance 
to probe the internal quark-gluon configurations of 
exotic-hadron candidates \cite{kks2013,cks2016}
in Sec.\,\ref{counting}. Then, the GPDs
for the nucleon and exotic-hadron candidates are discussed 
in Sec.\,\ref{gpds} including possible experimental studies at J-PARC
\cite{gpd-gda,J-PARC-GPD1,J-PARC-DY}.
Since there is no stable exotic hadron, the direct GPD studies 
do not seem to be possible for exotic hadrons, except for transition GPDs 
such as the ones from the nucleon to an exotic hadron.
In this respect, the GDAs could be appropriate 3D 
structure functions for studying exotic hadrons 
\cite{gpd-gda} because an exotic-hadron
pair could be produced in the final state even for an unstable
hadron. We explain such GDAs in Sec.\,\ref{gdas}, and it 
should be a possible project as two-photon processes at KEKB
\cite{gpd-gda,in-progress}.
We summarize our studies in Sec.\,\ref{summary}.

\section{Constituent-counting rule for exotic hadrons}
\label{counting}

\begin{wrapfigure}[11]{r}{0.41\textwidth}
   \vspace{-0.3cm}
   \begin{center}
     \includegraphics[width=4.7cm]{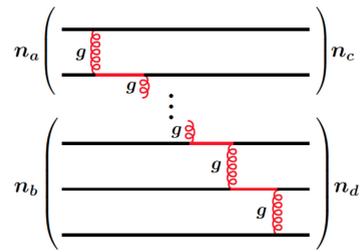}
   \end{center}
\vspace{-0.50cm}
\caption{A hard-gluon exchange process \cite{kks2013}.}
\label{fig:hard-gluon-exchange}
\vspace{-0.7cm}
\end{wrapfigure}

\begin{figure}[b]
\vspace{-0.3cm}
\begin{center}
\hspace{0.40cm}
\begin{minipage}{0.48\textwidth}
\hspace{0.0cm}
\vspace{0.0cm}
     \includegraphics[width=5.0cm]{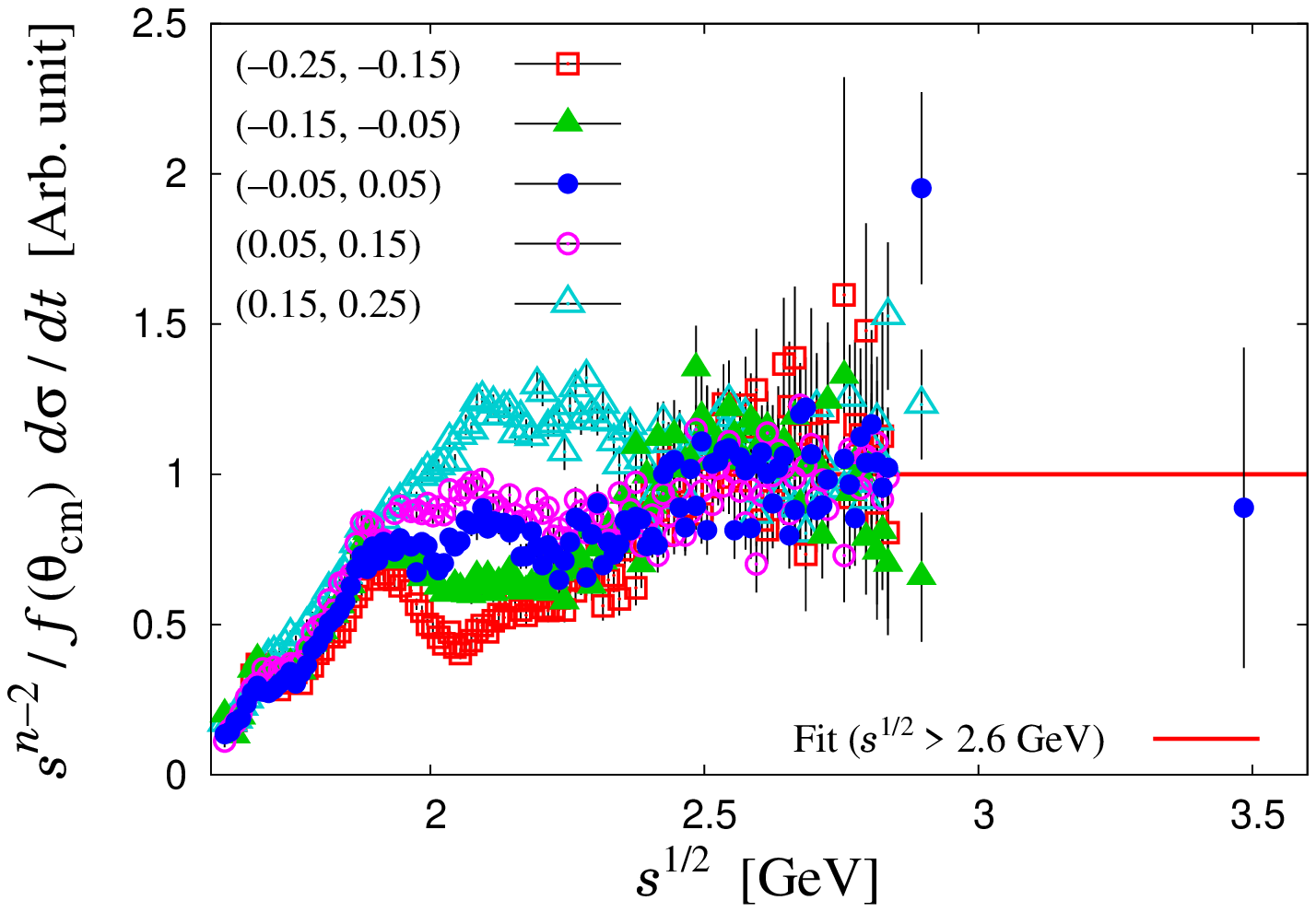}
\vspace{-0.45cm}
\caption{Scaling in $\gamma + p \to K^+ + \Lambda$
 \cite{cks2016}.}
\label{fig:gamma-lambda}
\end{minipage}
\vspace{-0.30cm}
\begin{minipage}{0.48\textwidth}
\hspace{0.3cm}
     \includegraphics[width=4.7cm]{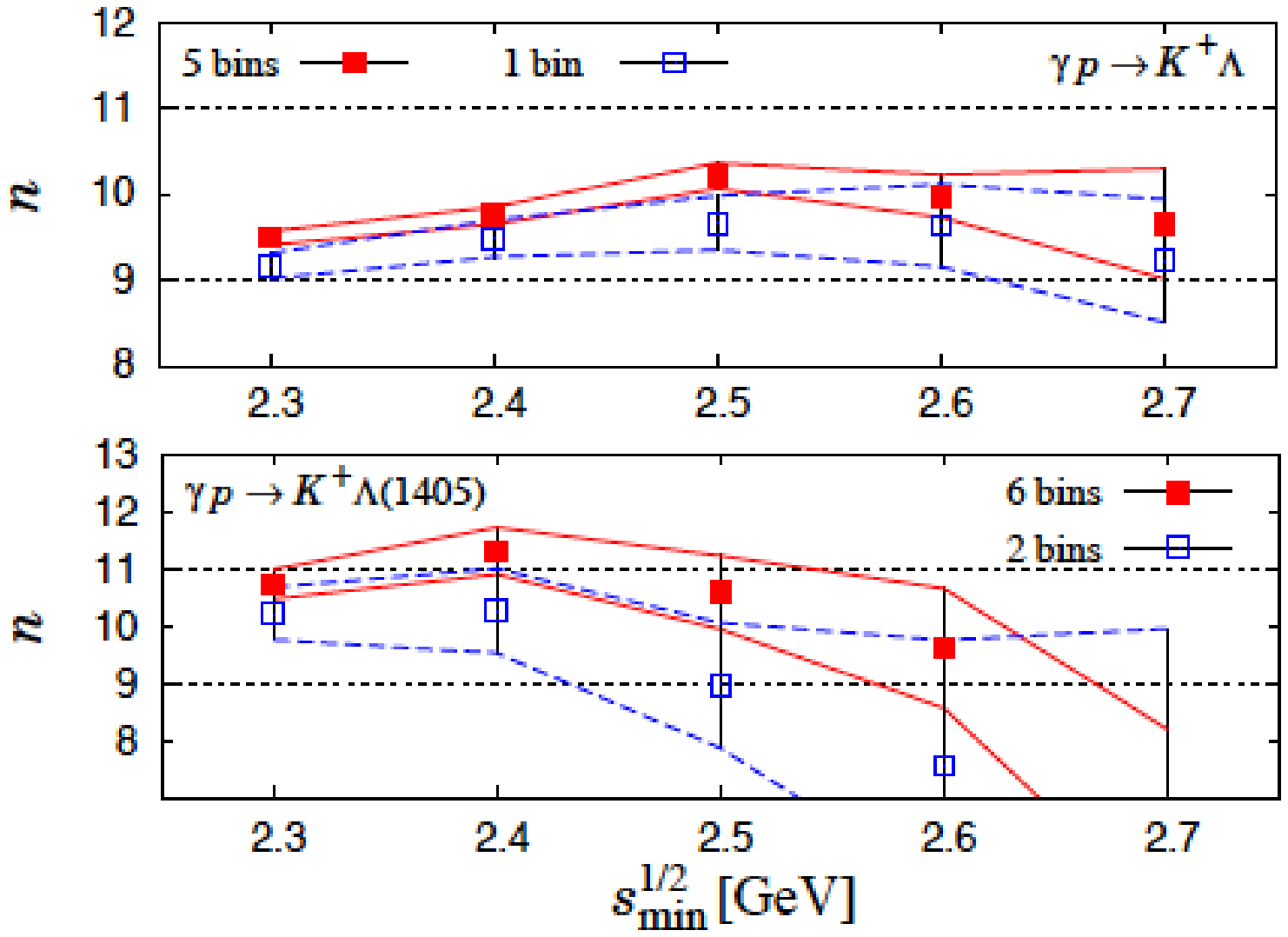}
\vspace{-0.30cm}
\caption{Energy dependence of factor $n$ \cite{cks2016}.}
\label{fig:factor-n}
\end{minipage} 
\end{center}
\vspace{-0.0cm}
\end{figure}

The constituent-counting rule was predicted in hard exclusive
reactions by perturbative QCD. As shown typically in 
Fig.\,\ref{fig:hard-gluon-exchange}, the two-body 
hadron reaction $a+b \to c+d$ with large $p_T$ should 
occur due to hard gluon exchanges between quarks in the hadrons.
By considering the hard gluon and quark propagators together
with other kinematical factors, we obtain the scaling relation
for the cross section as \cite{kks2013}
\begin{align}
\frac{d\sigma_{ab \to cd}}{dt} 
    = \frac{1}{s^{\, n-2}} \, f_{ab \to cd}(t/s),
\label{eqn:cross-counting}
\end{align}
where $s$ and $t$ are Mandelstam variables,
$f(t/s)$ is a function depends on the scattering angle, and
$n$ is the total number of constituents given by $n = n_a+n_b+n_c+n_d$.
The high-energy cross section is proportional to $1/s^{\, n-2}$ 
with the number of constituents ($n$), and it is called
the constituent-counting rule. 
It has been confirmed experimentally by various hard processes
in lepton-hadron and hadron-hadron exclusive reactions, even
for compound systems of hadrons such as the deuteron, $^3$H, and $^3$He.
Since it is not easy to judge whether or not discovered hadrons 
are actually exotic only by low-energy observables,
we think that it is appropriate to use high-energy reaction processes
to find internal constituents of exotic hadrons.
Especially, the appropriate degrees of freedom are quarks and gluons
at high energies, so that such approaches could be promising as
future exotic-hadron studies.

For example, we investigate an exotic-hadron candidate $\Lambda (1405)$,
which is considered to be $\bar K N$ molecule or pentaquark state
\cite{Lam-1405}, by the constituent counting rule. First, we look 
at the ground $\Lambda$ production $\gamma + p \to K^+ + \Lambda$, 
and the cross section scales as shown in Fig.\,\ref{fig:gamma-lambda}.
The scaling factor $n$ determined from the data depends the minimum
energy $\sqrt{s_{min}}$ for the scaling region as shown 
in Fig.\,\ref{fig:factor-n}. The factor $n=9 \ (=1+3+2+3)$ means that
the ground $\Lambda$ is consistent with the $qqq$ picture.

\begin{wrapfigure}[10]{r}{0.44\textwidth}
   \vspace{-0.25cm}
   \begin{center}
     \includegraphics[width=4.8cm]{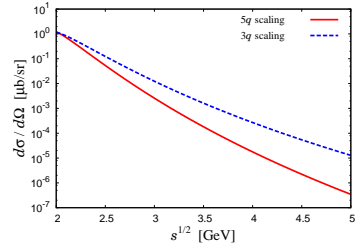}
   \end{center}
\vspace{-0.80cm}
\caption{$\pi p \to K^0 \Lambda (1405)$ cross sections
\cite{kks2013}.}
\label{fig:cross-section}
\vspace{-0.4cm}
\end{wrapfigure}

There are also data for the $\Lambda (1405)$ production
$\gamma + p \to K^+ + \Lambda (1405)$. We analyzed its data
and obtained the scaling factor $n$ as shown in Fig.\,\ref{fig:factor-n}.
The figure implies an interesting tendency that $\Lambda (1405)$ looks
like a five-quark state ($\bar K N$ molecule or pentaquark state) 
at relatively-low energies but it becomes three-quark one 
at high energies. However, the errors of $n$ are too large 
to draw a solid conclusion at this stage.
Future experimental studies including by a 12-GeV JLab experiment
are needed. Similar studies should be possible by hadron 
accelerator facilities such as J-PARC. As an example, we estimated 
the cross sections for $\pi +p \to K^0 +\Lambda (1405)$ by taking 
$n_{\Lambda (1405)}=3$ or $5$. There are large differences between
the curves, so that the internal structure of $\Lambda (1405)$
can be investigated through the counting rule at J-PARC.

\section{Generalized parton distributions and J-PARC project}
\label{gpds}

\begin{wrapfigure}[7]{r}{0.33\textwidth}
   \vspace{+0.1cm}
   \hspace{-0.2cm}
     \includegraphics[width=4.6cm]{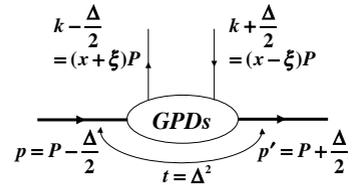}
\vspace{-0.45cm}
\caption{Kinematics of GPDs \cite{J-PARC-DY}.}
\label{fig:gpd-kinematics}
\vspace{-0.4cm}
\end{wrapfigure}

The GPDs are defined for the nucleon by off-forward nucleon matrix
elements of quark (and gluon) bilocal operators as
\cite{gpd-gda,J-PARC-GPD1,J-PARC-DY}
\begin{align}
& \! \! \! 
\int \! 
\frac{d y^-}{4\pi}e^{i x P^+ y^-}   \! \! 
 \left< p' \left| 
 \bar{q}(-y/2) \gamma^+ q(y/2) 
 \right| p \right> \Big |_{y^+ = \vec y_\perp =0}
 \! \! 
 =  \frac{1}{2  P^+} \, \bar{u} (p') 
 \left [ H^q (x,\xi,t) \gamma^+
     + E^q (x,\xi,t)  \frac{i \sigma^{+ \alpha} \Delta_\alpha}{2 \, m_N}
 \right ] u (p) ,
\nonumber \\
& \! \! \! 
\int \! 
\frac{d y^-}{4\pi}e^{i x P^+ y^-} \! \!
 \left< p' \left| 
 \bar{q}(-y/2) \gamma^+ \gamma_5 q(y/2) 
 \right| p \right> \Big |_{y^+ = \vec y_\perp =0}
 \! \! 
 =  \frac{1}{2  P^+} \, \bar{u} (p') 
 \left [ \tilde{H}^q (x,\xi,t) \gamma^+ \gamma_5
     + \tilde{E}^q (x,\xi,t)  \frac{\gamma_5 \Delta^+}{2 \, m_N}
 \right ] u (p) ,
\nonumber\\[-7pt]
\label{eqn:gpd}
\end{align}
\vspace{-0.6cm} \\
where $\sigma^{\alpha\beta}$ is 
$\sigma^{\alpha\beta}=(i/2)[\gamma^\alpha, \gamma^\beta]$,
gauge-link operators for the gauge invariance are not explicitly written, 
$H^q(x,\xi,t)$ and $E^q (x,\xi,t)$ are the unpolarized quark GPDs, 
and $\tilde{H}^q (x,\xi,t)$ and $\tilde{E}^q (x,\xi,t)$ are 
the polarized ones. 
By the momentum assignments in Fig.\,\ref{fig:gpd-kinematics},
the variables $x$ and $\xi$ are defined by
\vspace{-0.35cm}
\begin{align}
x & = \frac{\left( (k-\Delta/2)+(k+\Delta/2) \right)^+}{(p+p\,')^+}
=\frac{k^+}{P^+} , 
\nonumber \\
\xi & = \frac{(p-p\,')^+}{(p+p\,')^+} = \frac{-\Delta^+}{2P^+} 
= \frac{\left( (k-\Delta/2)-(k+\Delta/2) \right)^+}{(p+p\,')^+},
\label{eqn:x-xi}
\end{align}
so that $x$ and $2\xi$ are the light-cone momentum fractions 
of the average momentum and momentum transfer for the relevant quarks
to the average momentum of the parent nucleon. The range of $x$ is
from $-1$ to $1$, whereas $\xi$ is from $0$ to $1$.
The GPDs have necessary information to find 
3D structure of the nucleon, and they are
key quantities in the studies of hadron tomography.
There are three important features in the GPDs.
(I) 
In the forward limit ($\Delta\to 0$, $\xi\to 0$ and $t \rightarrow 0$),
they become unpolarized and longitudinally-polarized 
parton distribution functions (PDFs) for the nucleon:
$ H^q (x, 0, 0) = q(x)$ and $\tilde{H}^q (x, 0, 0) = \Delta q(x)$.
(I\hspace{-.1em}I)
The first moments of the unpolarized GPDs are
Dirac and Pauli form factors of the nucleon, and those of
the polarized GPDs are axial and pseudoscalar form factors:
\begin{align}
\! \! \! \!
\int_{-1}^{1} \! \! \! dx \, H^q(x,\xi,t)  = \! F_1^q (t), 
\int_{-1}^{1} \! \! \! dx \, E^q(x,\xi,t) = \! F_2^q (t),
\int_{-1}^{1} \! \! \! dx \, \tilde{H}^q(x,\xi,t)  = \! g_A^q (t), 
\int_{-1}^{1} \! \! \! dx \, \tilde{E}^q(x,\xi,t) = \! g_P^q (t).
\label{sumrule}
\end{align}
(I\hspace{-.1em}I\hspace{-.1em}I)
The second moment of a GPD combination becomes the quark total
angular momentum:
$J^q = \int_{-1}^{1} dx \, x \, [ H^q (x,\xi,t=0) +E^q (x,\xi,t=0) ] / 2$,
which makes it possible to find the the quark orbital-angular-momentum 
contribution to the nucleon spin by $J^q = \Delta q /2 + L^q$.
We now explain how the GPDs can be investigated at hadron facilities
such as J-PARC in addition to the usual deeply virtual Compton scattering
(DVCS) in the following subsections.

\subsection{Generalized parton distributions in $p + p \to N + \pi +B$}
\label{2-3-process}

\begin{wrapfigure}[9]{r}{0.37\textwidth}
   \vspace{-0.1cm}
\hspace{0.0cm}
     \includegraphics[width=5.0cm]{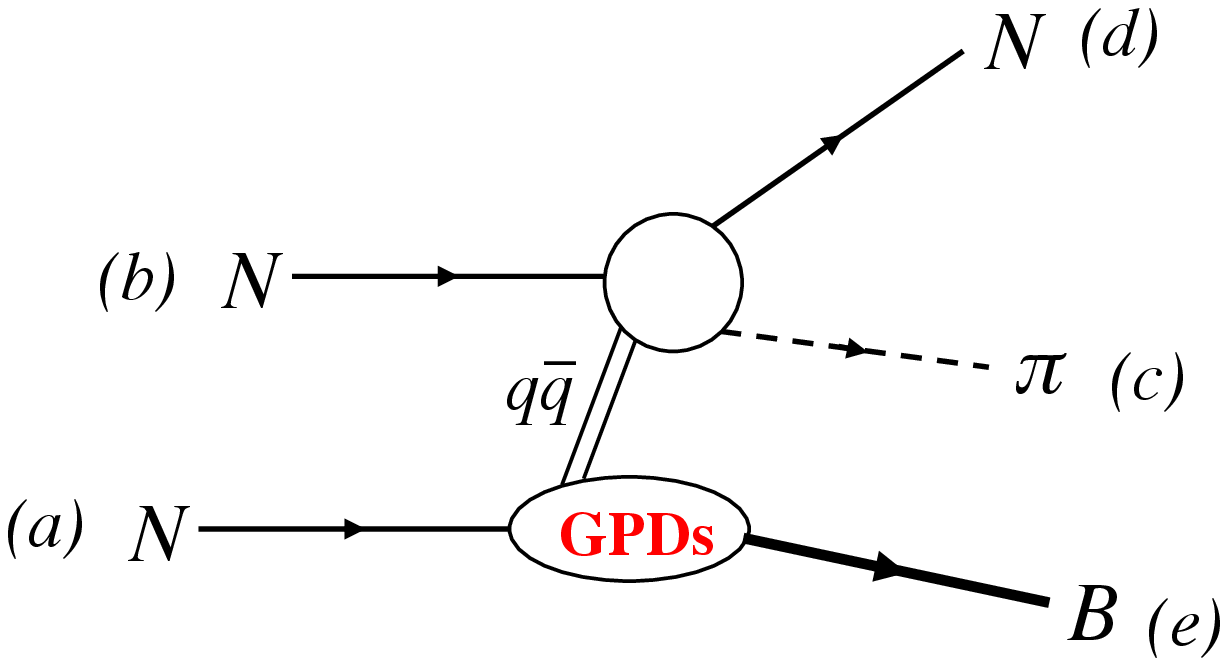}
\vspace{-0.20cm}
\caption{$NN \to N\pi B$ process for GPDs.}
\label{fig:nn-npib-gpds}
\vspace{-0.7cm}
\end{wrapfigure}

\begin{figure}[b!]
\vspace{-0.2cm}
\begin{center}
\hspace{0.0cm}
\begin{minipage}{0.60\textwidth}
     \includegraphics[width=8.2cm]{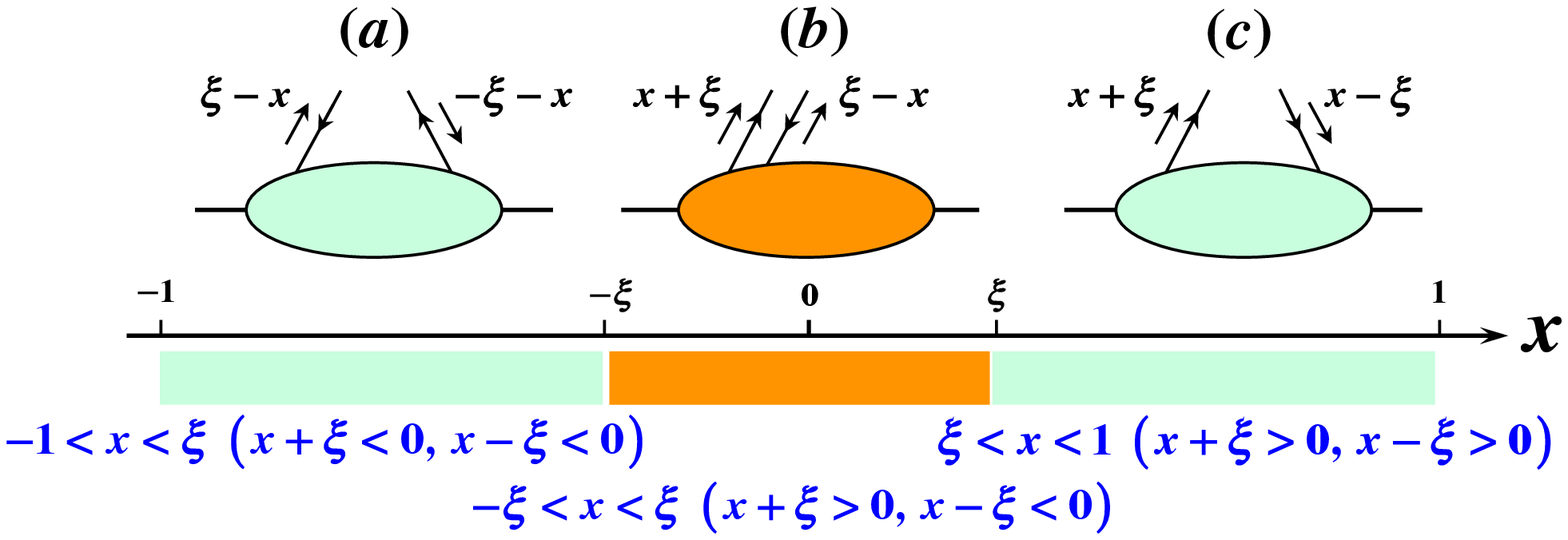}
\vspace{-0.3cm}
\caption{Three kinematical regions of GPDs:
$(a)$ emission and reabsorption of an antiquark,
$(b)$ emission and a quark and an antiquark,
$(c)$ emission and reabsorption of a quark.}
\label{fig:egpd-rbl}
\end{minipage}
\hspace{0.3cm}
\vspace{-0.4cm}
\begin{minipage}{0.36\textwidth}
\vspace{0.1cm}
\hspace{0.4cm}
     \includegraphics[width=4.5cm]{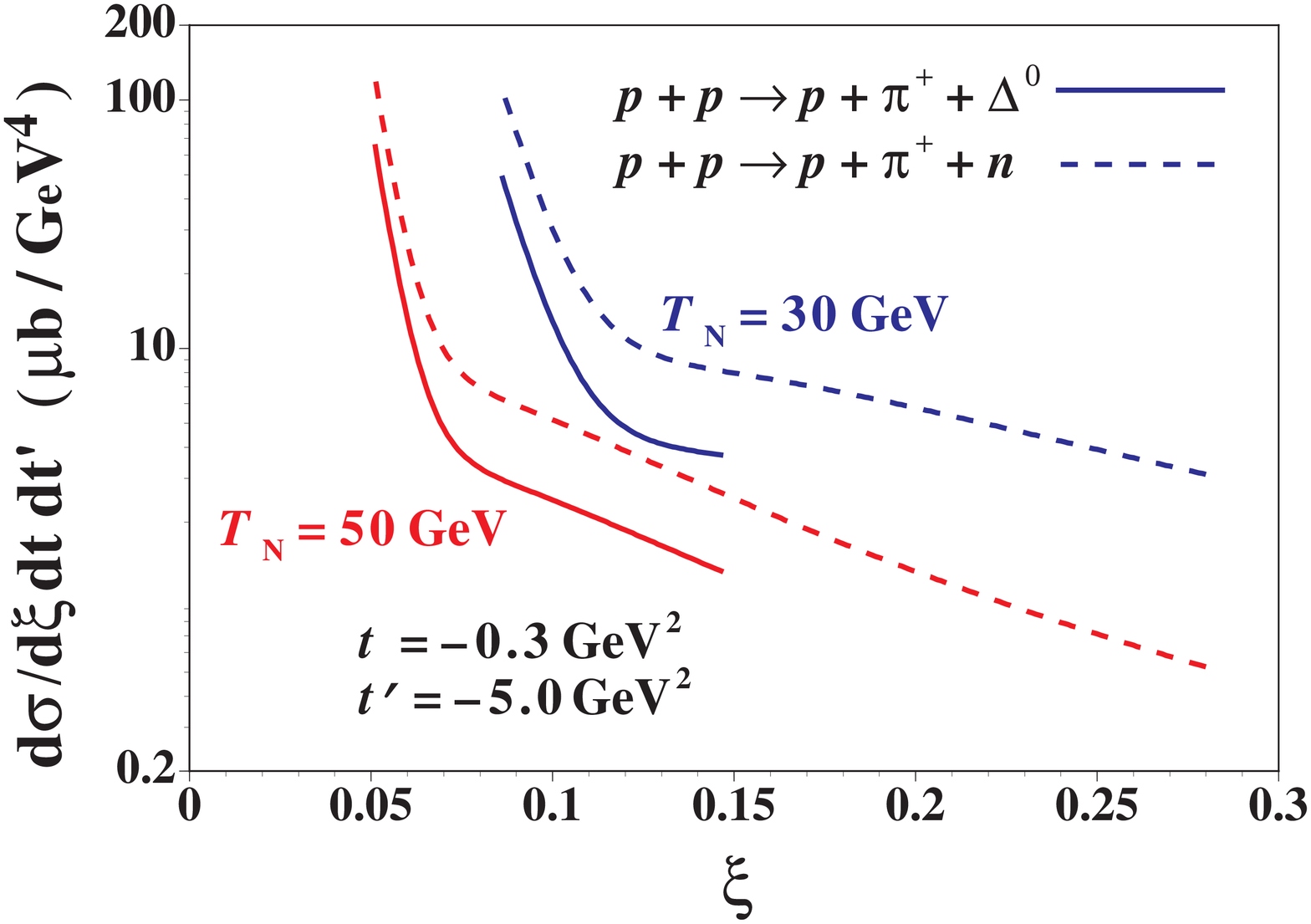}
\vspace{-0.3cm}
\caption{Cross sections as the function of $\xi$ 
for J-PARC kinematics \cite{J-PARC-GPD1}.}
\label{fig:cross-xi}
\end{minipage} 
\end{center}
\end{figure}

The GPDs have been experimentally investigated mainly in the DVCS.
However, it is possible to study them in hadron reactions.
Hadron-reaction rates are generally larger than the electromagnetic
Compton scattering, so that extended kinematical regions could be probed. 
Furthermore, it is a unique opportunity that the interesting 
Efremov-Radyushkin-Brodsky-Lepage (ERBL) region
can be studied in the reaction $N+N \to N + \pi + B$
of Fig.\,\ref{fig:nn-npib-gpds} \cite{J-PARC-GPD1}.
There are three kinematical regions in the GPDs as shown in 
Fig.\,\ref{fig:egpd-rbl}. The regions $(a)$ and $(c)$ are
the DGLAP (Dokshitzer-Gribov-Lipatov-Altarelli-Parisi) regions,
and their GPDs corresponds to the antiquark and quark distributions, 
respectively. The $(b)$ is the ERBL region, and the GPDs
of this region indicate quark-antiquark distributions.

We define the Mandelstam variables as
$s' = (p_c+p_d)^2$, $t'=  (p_{b}-p_{d})^2$, and $u' = (p_b-p_c)^2$.
For investigating the GPDs, we consider the hard region
$s', |t'|, |u'| \gg m_N^2$.
In the reaction $N+N \to N + \pi + B$ of Fig.\,\ref{fig:nn-npib-gpds}, 
it is possible to probe the ERBL region.
We estimated its cross sections for J-PARC kinematics
of 30 GeV and 50 GeV proton-beam energies
in Fig.\,\ref{fig:cross-xi} by taking $t=(p_a-p_e)^2=-0.3$ GeV$^2$ 
and $t'=-5.0$ GeV$^2$ as an example \cite{J-PARC-GPD1}.
Here, the neutron or $\Delta^0$ are assumed
for the final-state baryon $B$, and the pion- and $\rho$-pole terms
are used for the ERBL GPDs. These cross-section estimates are 
intended for proposing a future J-PARC experiment.
The GPDs, $H$, $E$, $\tilde H$, and $\tilde E$, for the nucleon
and the $N \to \Delta$ transition are involved in the cross section,
and they are theoretically estimated by the meson-pole contributions.
The interactions for $\pi N \to \pi N$ and $\rho N \to \pi N$,
which correspond to the upper part of Fig.\,\ref{fig:nn-npib-gpds},
are determined from the measurements of the BNL-E838 experiment
\cite{J-PARC-GPD1}.
The hadron facility measurements at J-PARC and possibly at GSI-FAIR
should provide new information on the GPDs in the unique ERBL region.
In order to determine the GPDs for the studies of the nucleon spin
and 3D structure, all the kinematical regions
should be investigated, so that the J-PARC measurement is
valuable and complementary to the usual DVCS experiments.

\subsection{Generalized parton distributions in exclusive Drell-Yan}
\label{exclusive-dy}

\begin{wrapfigure}[10]{r}{0.37\textwidth}
   \vspace{+0.2cm}
\hspace{0.0cm}
     \includegraphics[width=5.0cm]{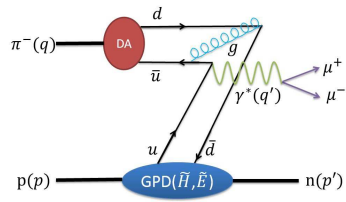}
\vspace{-0.30cm}
\caption{GPD contribution to exclusive Drell-Yan \cite{J-PARC-DY}.}
\label{fig:ExclusiveDY}
\vspace{-0.7cm}
\end{wrapfigure}

There is another possibility to investigate the GPDs at J-PARC
by using an exclusive Drell-Yan process $\pi^- + p \to \mu^+\mu^- + n$
\cite{J-PARC-DY} because high-momentum unseparated-hadron 
(essentially pion) beamline will become available soon. 
A typical GPD process for the exclusive
dimuon production is shown in Fig.\,\ref{fig:ExclusiveDY},
where the process is described by the GPDs combined with
the pion distribution amplitude (DA). However, the distribution amplitude 
should be constrained by other measurements such as on the $\gamma \to \pi$
transition form factor of the Belle and BaBar experiments, 
although there are still some ambiguities whether it is the asymptotic form, 
the Chernyak-Zhitnitsky one, or something in between. 

\begin{figure}[h!]
\vspace{-0.3cm}
\begin{center}
\hspace{0.0cm}
\begin{minipage}{0.49\textwidth}
\hspace{0.8cm}
     \includegraphics[width=5.2cm]{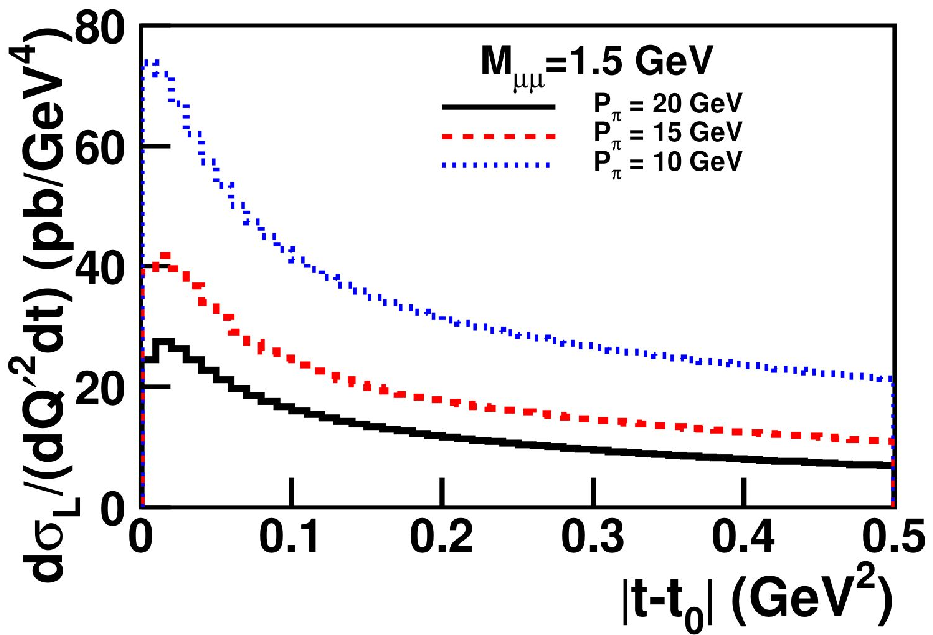}
\vspace{-0.35cm}
\caption{$t$ dependence of $\pi^- + p \to \mu^+\mu^- + n$ cross sections
for J-PARC kinematics \cite{J-PARC-DY}.}
\label{fig:exclusive-dy-jparc-1}
\end{minipage}
\hspace{0.3cm}
\vspace{-0.0cm}
\begin{minipage}{0.47\textwidth}
\vspace{0.3cm}
\hspace{1.2cm}
     \includegraphics[width=3.7cm]{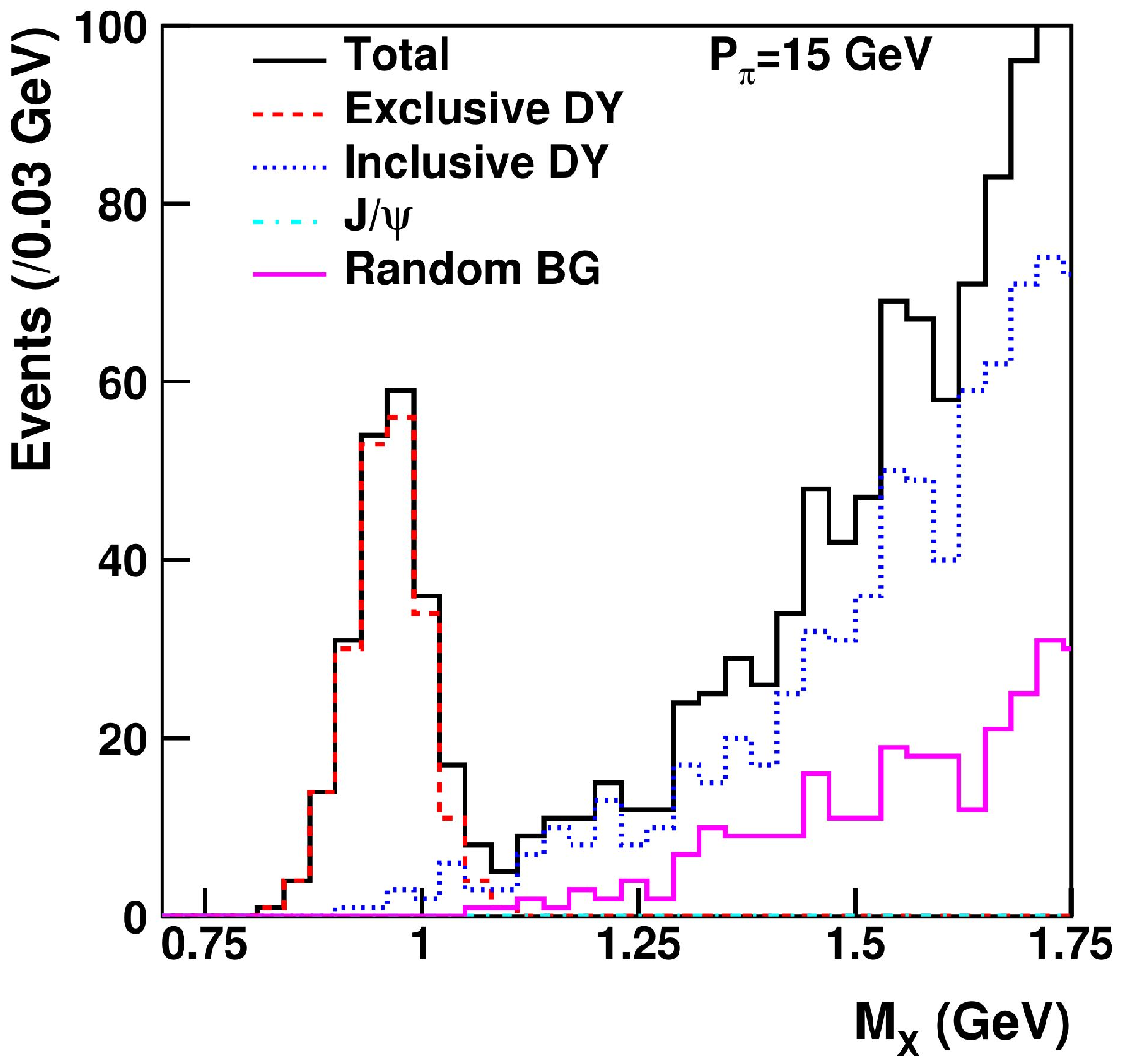}
\vspace{-0.6cm}
\caption{Simulations of missing-mass ($M_X$) spectra of $\mu^+\mu^-$ 
events \cite{J-PARC-DY}.}
\label{fig:exclusive-dy-jparc-2}
\end{minipage} 
\end{center}
\vspace{-0.9cm}
\end{figure}

The GPDs $\tilde H$ and $\tilde E$ contribute to the cross section,
and a typical GPD parametrization GK2013 is used for showing
numerical results in Fig.\,\ref{fig:exclusive-dy-jparc-1},
where three possible pion momenta, $p_\pi=10$, 15, and 20 GeV,
are taken and $t_0$ is the limiting value of the Mandelstam variable $t$
at the zero scattering angle. This estimate is intended for 
a J-PARC proposal, and actual GPDs should be determined from future 
measurements at J-PARC together with other data including DVCS ones.
Next, Monte Carlo simulations of missing-mass ($M_X$) spectra of $\mu^+\mu^-$ 
events are shown in Fig.\,\ref{fig:exclusive-dy-jparc-2}.
There are contributions from various sources: exclusive Drell-Yan (red), 
inclusive Drell-Yan (blue), J/$\psi$ (green), and random background (purple).
It is obvious that the exclusive Drell-Yan dominates the events at
$M_X \simeq 1$ GeV, so that it is a feasible experiment at J-PARC.
At JLab, there is a proposed experiment to measure the opposite
process $\gamma^* N \to \pi N$; however, it is a timelike photon
process at J-PARC and the probed $x$ region is smaller than the JLab case
\cite{J-PARC-DY}, so that the J-PARC experiment is a valuable
complementary contribution to the GPD studies.

\vspace{-0.1cm}

\subsection{Generalized parton distributions for exotic hadrons}
\label{exotic-gpds}

In addition to the constituent counting rule in Sec.\,\ref{counting},
the GPDs, in general the 3D structure functions,
can be used for exotic-hadron studies \cite{gpd-gda}
because the relevant degrees of freedom are quarks and gluons 
at high energies and because exotic signatures should 
appear in the 3D structure.
The GPDs have two types of information on internal structure of
a hadron: (1) longitudinal-momentum distributions, namely PDFs 
($f_n (x)$), (2) transverse form factors ($F_n^h (t, x) $) which
also depend on the longitudinal-momentum fraction $x$.
Therefore, the GPDs could be expressed by their multiplication
as a simple form:
\begin{align}
H_q^h (x,\xi=0,t)= f_n (x) \, F_n^h (t, x) ,
\label{eqn:gpd-paramet1}
\end{align}
where $n$ indicates the number of valence quarks.

\begin{figure}[h!]
\vspace{+0.1cm}
\begin{center}
\hspace{0.0cm}
\begin{minipage}{0.48\textwidth}
\hspace{0.8cm}
     \includegraphics[width=5.0cm]{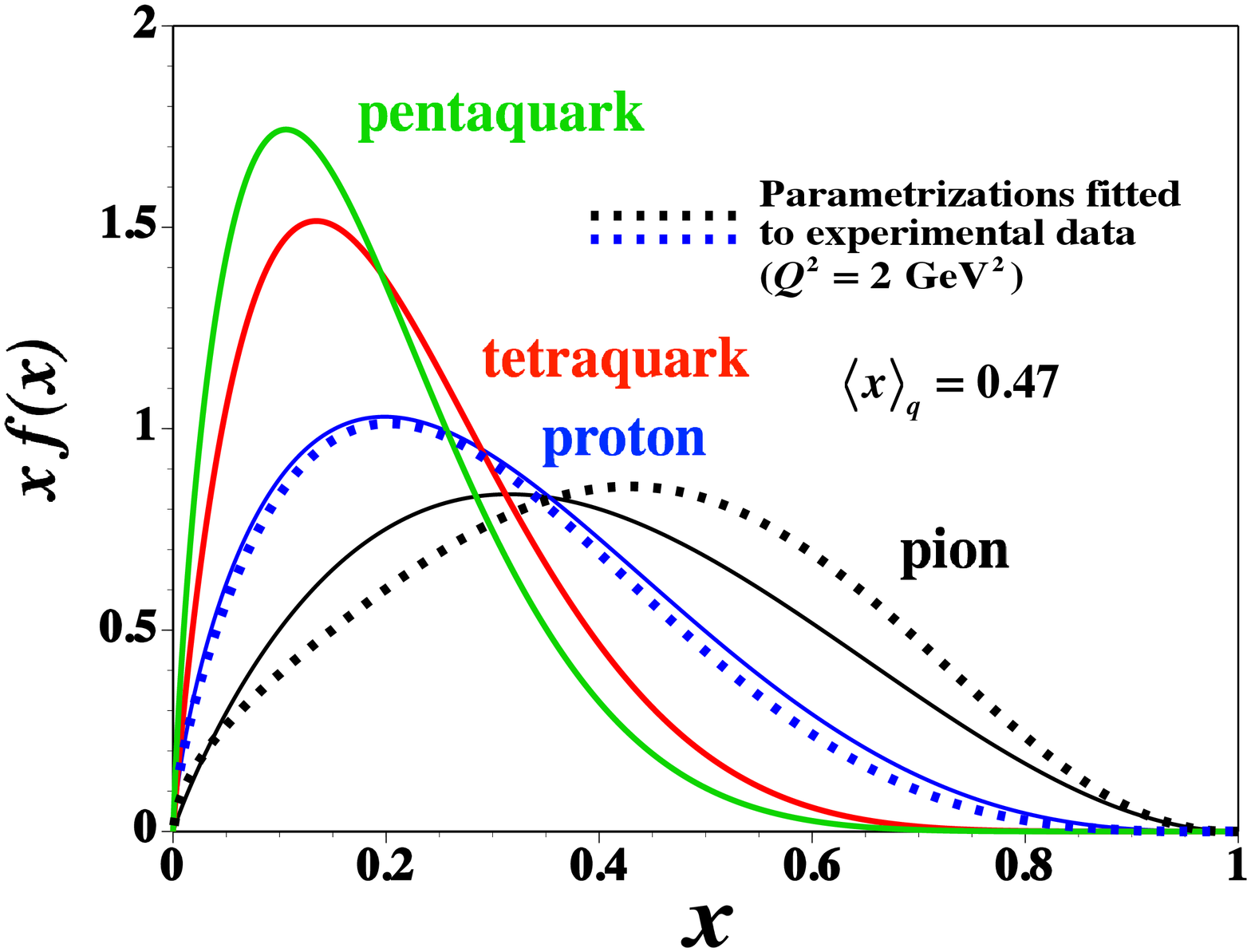}
\vspace{-0.33cm}
\caption{Valence-quark distributions for pion, proton,
tetraquark, and pentaquark \cite{gpd-gda}.}
\label{fig:exotic-pdf}
\end{minipage}
\hspace{0.3cm}
\begin{minipage}{0.48\textwidth}
\vspace{-0.3cm}
\hspace{0.8cm}
     \includegraphics[width=5.0cm]{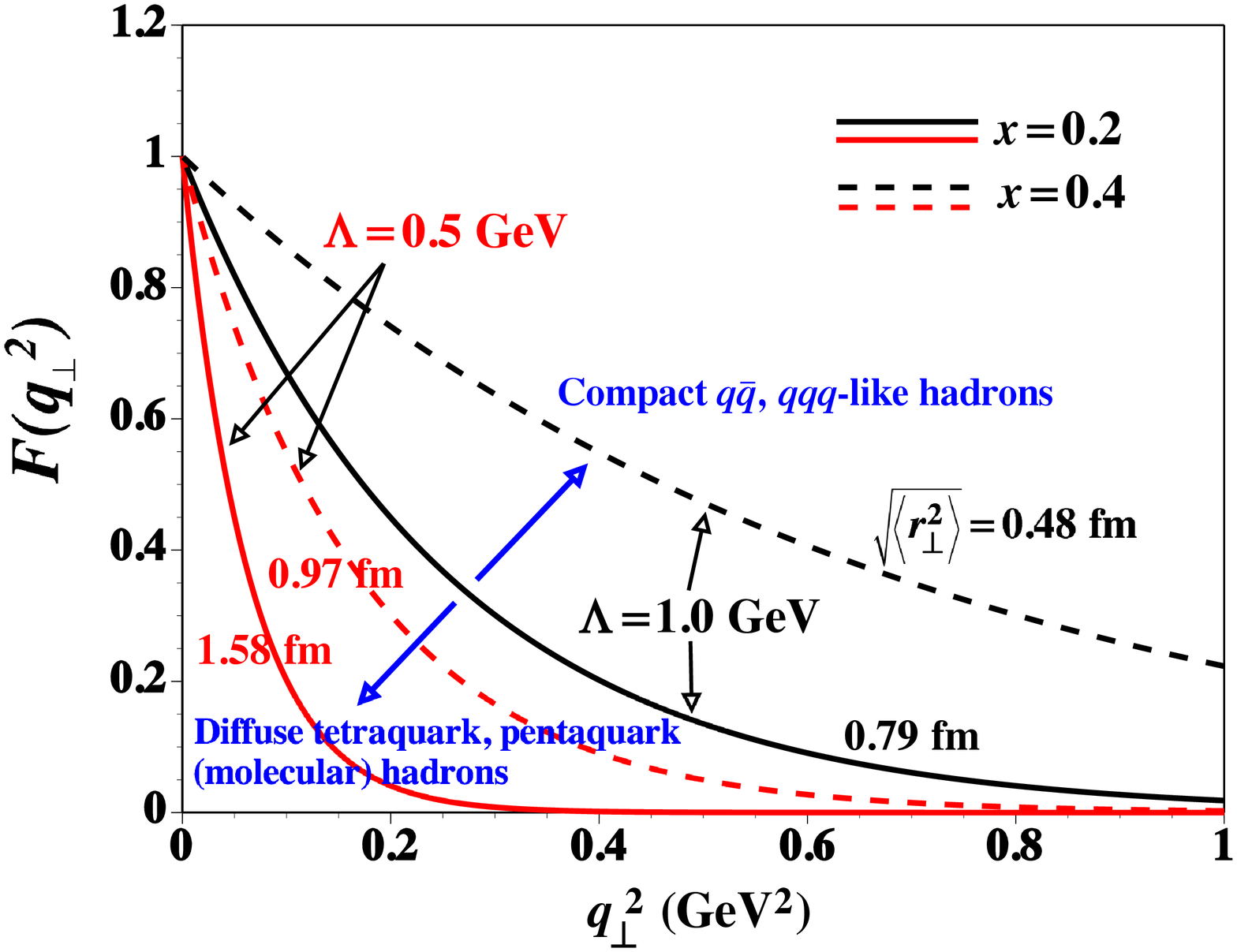}
\vspace{-0.3cm}
\caption{Transverse form factors at $x=0.2$, 0.4 \cite{gpd-gda}.}
\label{fig:transverse-gpd}
\end{minipage} 
\end{center}
\vspace{-0.9cm}
\end{figure}

A simple $x$-dependent form of the longitudinal PDFs is
$ f_n (x) = C_n \, x^{\alpha_n} \, (1-x)^{\beta_n} $.
We consider valence-quark distributions in this form.
Then, if $\beta_n$ is known, the parameters $C_n$ and $\alpha_n$ 
are determined by the valence-quark number
$ \int_0^1 dx \, f_n (x) = n$, and the quark momentum
$ \int_0^1 dx \, x \, f_n (x) = \langle x \rangle_q$.
The parameter $\beta_n$ is associated with the functional behavior 
in the elastic limit $x \to 1$, and it is determined by 
the constituent counting rule as $\beta_n = 2n -3+2\Delta S_z$ 
with the spin factor $\Delta S_z=|S_z^q-S_z^h|$. 
From these three conditions, we theoretically obtain the PDFs
not only for the ordinary hadrons, such as pion and nucleon,
but also for exotic tetraquark and pentaquark hadrons,
and the results are shown by the solid curves 
in Fig.\,\ref{fig:exotic-pdf}.
In comparison, typical PDFs of the pion and the proton extracted from
experimental measurements are shown by the dotted curves
at $Q^2=2$ GeV$^2$.
Our theoretical curves and the PDF parametrizations agree 
with each other. Therefore, the predications for the new PDFs
of the tetraquark and pentaquark hadrons should be interesting
if they are measured by experiments.

Next, we explain that the transverse form factor $F_n^h (t, x)$ also
has information on exotic nature \cite{gpd-gda}.
A simple form of this function is an exponential form
$F_n^h (t,x) = e^{(1-x) t/(x \Lambda^2)}$,
where $\Lambda$ is cutoff parameter for the transverse momentum
and it is related to the 
the root-mean-square radius by
$ \langle  r_\perp^2 \rangle = 4(1-x) / (x \Lambda^2 )$.
Since the pion and nucleon form factors are roughly given by
the monopole and dipole forms ($1/(|t|+\Lambda^2)$, $1/(|t|+\Lambda^2)^2$),
the Gaussian form could be too steep as a function of $q_\perp^2$
and it is realized only in light nuclei. 
We do not step into such details in this article.
The results are shown in Fig.\,\ref{fig:transverse-gpd}
for $x=0.2$ and 0.4. Depending on compact $q\bar q$/$qqq$ hadrons
or diffuse molecular exotic hadrons, the transverse form factors
are different, so it should be an appropriate quantity to find
the internal structure of exotic-hadron candidates.

\begin{wrapfigure}[9]{r}{0.37\textwidth}
   \vspace{+0.2cm}
\hspace{0.3cm}
     \includegraphics[width=4.2cm]{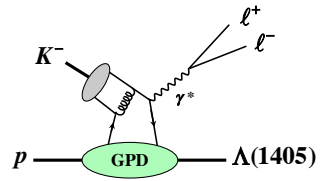}
\vspace{-0.20cm}
\caption{Transition GPDs at J-PARC.}
\label{fig:transition-gpds}
\end{wrapfigure}
\vspace{0.2cm}

The GPDs of the exotic hadrons could not be observed directly
at least at this stage because the exotic hadrons cannot exist 
as stable fixed targets. Of course, transition GPDs such as
for proton $\to$ an exotic hadron can be investigated in future.
For example, by using high-momentum kaon beam which will 
become available at J-PARC, it is possible to investigate
the transition GPDs of $p \to \Lambda (1405)$ 
as shown in Fig.\,\ref{fig:transition-gpds}, and they 
should reflect the exotic nature of $\Lambda (1405)$.
Since the transition GPDs, especially on exotic hadrons,
are not well studied, theoretical and experimental efforts
are needed to push this kind of experimental projects
in future.

\vfill\eject

\section{Generalized distribution amplitudes and KEKB project}
\label{gdas}

\begin{wrapfigure}[12]{r}{0.38\textwidth}
   \vspace{+0.3cm}
\hspace{0.3cm}
     \includegraphics[width=4.5cm]{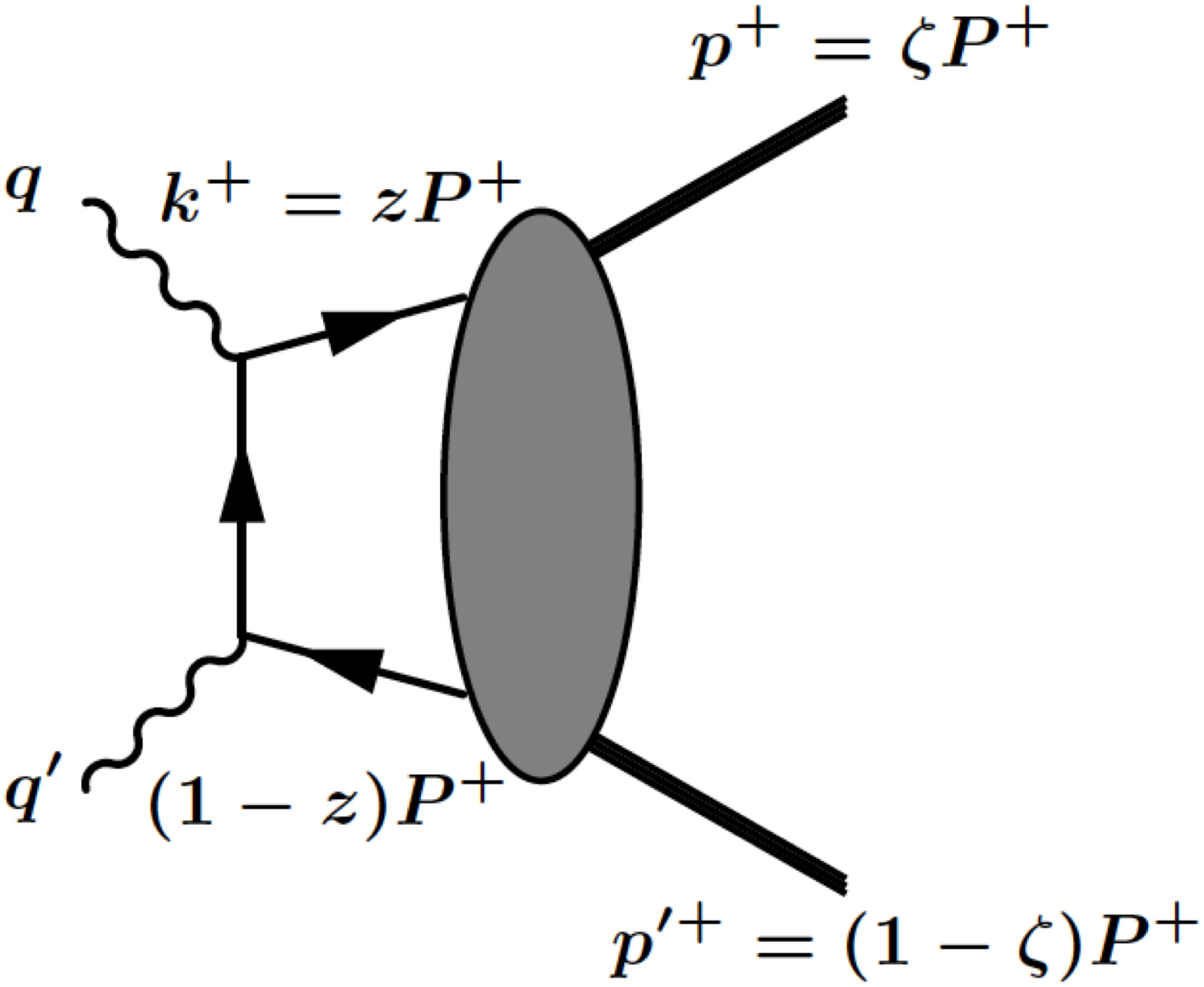}
\vspace{-0.30cm}
\caption{$\gamma\gamma \to h \bar h$ and GDAs \cite{gpd-gda}.}
\label{fig:dga-1}
\end{wrapfigure}

It is an interesting idea to use the transition GPDs 
for exotic-hadron studies; however, another promising way 
of studying 3D structure functions for exotic hadrons
is to use general distribution amplitudes (GDAs) because 
exotic hadrons can be produced in s-channel processes
\cite{gpd-gda}.
A two-photon process to produce an exotic-hadron pair 
$\gamma\gamma^* \to h \bar h$ is shown in Fig.\,\ref{fig:dga-1}.
If the kinematical condition, 
$Q^2 \, (=-q^2) \gg W^2 \, (=(p+p')^2) , \, \Lambda^2$,
is satisfied, we have the factorization of the process 
in terms of the hard photon interaction and the GDAs.
This two photon process is the $s$-$t$ crossed one to the DVCS
($\gamma^* h \to \gamma h$) where the GPDs are investigated,
so that the $s$-$t$ crossed 3D structure functions 
of the GPDs are the GDAs.

The quark GDAs are defined by the same lightcone operator
to the GPDs between the vacuum and the hadron pair $h\bar h$:
\begin{align}
\Phi_q^{h\bar h} (z,\zeta,s) 
= \int \frac{d y^-}{2\pi}\, e^{i (2z-1)\, P^+ y^-}
   \langle \, h(p) \, \bar h(p') \, | \, 
 \overline{\psi}(-y/2) \gamma^+ \psi(y/2) 
  \, | \, 0 \rangle \Big |_{y^+=\vec y_\perp =0} \, ,
\end{align}
and the gluon GDA is defined in the similar way.
The $s$-$t$ crossing relates the GDAs to GPDs,
and the relation is given by
\begin{align}
\Phi_q^{h\bar h} (z,\zeta,W^2) 
\longleftrightarrow
H_q^h \left ( x=\frac{1-2z}{1-2\zeta},
            \xi=\frac{1}{1-2\zeta}, t=W^2 \right ) .
\label{eqn:gda-gpd-relation}
\end{align}
It means that the GDAs contain the information 
of GPDs not only  in the usual physical region but also
in the unphysical regions:
$ 0 \le |x| < \infty$, $ 0 \le |\xi| < \infty$, 
$ |x| \le |\xi|$, and $ t \ge 0 $.
A simple functional form is given by
\begin{align}
\Phi_q^{h\bar h (I=0)} (z,\zeta,W^2) 
= N_{h(q)} \, z^\alpha (1-z)^\beta (2z-1) \, \zeta (1-\zeta) \, F_{h(q)} (W^2) ,
\label{eqn:gda-paramet}
\end{align}
where $F_{h(q)} (W^2)$ is a form factor of the quark part
of the energy-momentum tensor. According to the constituent counting rule,
it is expressed as 
$ F_{h(q)} (W^2) = 1 / [ 1 + (W^2-4 m_h^2)/\Lambda^2 ]^{n-1} $,
where the factor $n$ is $n=2$ for ordinary $q\bar q$ mesons and $n=4$ 
for tetraquark hadrons.

\begin{wrapfigure}[12]{r}{0.46\textwidth}
   \vspace{0.2cm}
\hspace{0.5cm}
     \includegraphics[width=5.0cm]{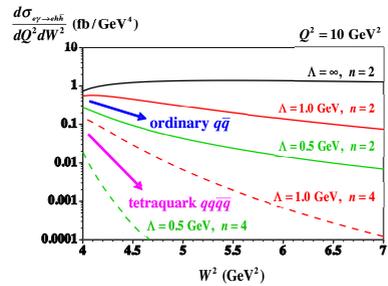}
\vspace{-0.30cm}
\caption{Cross section for $h\bar h$-pair production \cite{gpd-gda}.}
\label{fig:two-photon}
\end{wrapfigure}

The $f_0 (980)$ and $a_0 (980)$ mesons have been controversial 
hadrons which seem to be tetra-quark states or $K\bar K$ molecules
\cite{f0-a0}. Here, we show $W^2$ dependence of the cross section 
$e \gamma \to e' h \bar h$ in Fig. \ref{fig:two-photon}
for $h=f_0 (980)$ or $a_0 (980)$ by using the simple GDAs 
in Eq.\, (\ref{eqn:gda-paramet})
with the form factor suggested by the constituent counting rule.
There are large differences in the cross sections 
depending on the constituent number $n$,
$n=2$ (ordinary meson, $q\bar q$) or 
$n=4$ (tetraquark type, $qq\bar q \bar q$).
Even for the $n=4$ hadron, it is possible 
distinguish a compact $qq\bar q \bar q$ state
from a diffuse molecular state from the measurement.
Therefore, the GDAs should provide valuable information 
on internal structure of exotic hadrons, and such two-photon
physics is certainly a possible project at KEKB
\cite{in-progress}.

\section{Summary}
\label{summary}

It became crucial to investigate the 3D structure functions
for finding the original of nucleon spin including partonic 
orbital-angular-momentum contributions. One of such 3D structure functions
is the GPDs which have been investigated by the DVCS so far. However,
it is also possible to investigate them at hadron facilities such as J-PARC,
for example, by using the $2 \to 3$ process $p+p \to N + \pi +B$ and
the exclusive Drell-Yan process $\pi^- +p \to \mu^+ \mu^- +n$. 
In addition, the 3D structure functions, the GPDs and GDAs, can be
used for finding internal structure of exotic hadrons. In particular,
the GDAs should be appropriate 3D quantities for the exotic-hadron
studies because the exotic hadrons can be created in the final state, 
whereas it is difficult to measure the exotic-hadron GPDs directly 
because unstable exotic hadrons cannot be used as stable fixed targets.
The GDAs can be investigated by the two-photon processes, for example, 
at KEKB and it is a new promising direction of exotic-hadron studies.

\vspace{-0.0cm}
\section*{Acknowledgements}
This work was supported by Japan Society for the Promotion of Science (JSPS)
Grants-in-Aid for Scientific Research (KAKENHI) Grant
No. JP25105010.



\end{document}